\documentclass[twocolumn,prl,showpacs,preprintnumbers,amsmath,amssymb,superscriptaddress,unsortedaddress]{revtex4}

\usepackage{graphicx}
\usepackage{dcolumn}
\usepackage{bm}

\def \half{\frac{1}{2}}
\def \ngb{$n_{gb}$}
\def \nge{$n_{ge}$}
\def \Vds{$V_{ds}$}
\def \ngbHalf{$n_{gb}=\half$}
\def \ngeHalf{$n_{ge}=\half$}
\def \VdsZero{$V_{ds}=0$}

\def \Vgb{$V_{gb}$}
\def \Cc{$C_{c}$}

\begin{document}

\preprint{APS/123-QED}

\title{SET Backaction on the Single Electron Box}%

\author{B. A. Turek}
\author{K. W. Lehnert}
\author{A. Clerk}
\affiliation{Department of Applied Physics and Department of Physics, Yale University,
New Haven, CT 06511 USA}
\author{D. Gunnarsson}
\author{K. Bladh}
\author{P. Delsing}
\affiliation{Microtechnology Center at Chalmers MC2, Department of Microelectronics and Nanoscience,\\
Chalmers University of Technology and Goteborg University, SE-412 96, Goteborg, Sweden}
\author{R. J. Schoelkopf}
\affiliation{Department of Applied Physics and Department of Physics, Yale University,
New Haven, CT 06511 USA}%

\date{\today}

\begin{abstract}
We report an experimental observation of the backaction of a Single Electron Transistor (SET) measuring the Coulomb staircase of a single electron box.  As current flows through the SET, the charge state of the SET island fluctuates. These fluctuations capacitively couple to the box and cause changes in the position, width, and asymmetry of the Coulomb staircase. A sequential tunnelling model accurately recreates these effects, confirming this mechanism of the backaction of an SET. This is a first step towards understanding the effects of quantum measurement on solid state qubits.

\end{abstract}


\maketitle

In the recent work towards the goal of quantum computing, and in the study of single quantum systems in general, the Single Electron Transistor (SET) is often used as a measurement device. It has been proposed as a readout device for mechanical\cite{Blencowe}, spin\cite{Kane}, and charge \cite{Bouchiat} quantum systems, and has been successfully used to measure superconducting charge qubits \cite{LehnertT1}.  As with any amplifier, the SET must produce electrical noise on its input, perturbing the measured system and causing the unavoidable backaction of a quantum measurement.

SET backaction on a two level system has been studied extensively in the theoretical literature. It has been determined that the SET should be able to approach the quantum limit of backaction, where it dephases a qubit as rapidly as it is reads the qubit state\cite{Devoret}. Spectral components of the SET backaction at the two level system transition frequency can also contribute to transitions between two qubit states\cite{Aassime, Johansson}. A qubit could thus form a spectrum analyzer capable of probing previously inaccessible frequencies \cite{Schoelkopf}. These theoretical analyses presume SET backaction results from fluctuations in the charge state of the SET island caused by the drain-source current, but no experimental measurements exist confirming that this is the dominant or the sole mechanism of the SET's backaction. Indeed, it often appears that the SET can poison the Cooper-pair box, inducing non-equilibrium quasiparticles through other mechanisms\cite{TurekASC,MannikPRL}.

As a first quantitative test of SET backaction, we consider the SET and box operated in the normal (non-superconducting) state, created with the application of a 1 Tesla magnetic field. Analysis of the normal box is simpler than in the superconducting state because the box is no longer sensitive to parity and quasiparticle generation.  The normal SET can also be simply described by a sequential tunnelling model, which avoids the complication of the many possible quasiparticle-pair tunnelling cycles\cite{Pohlen} in the superconducting SET. Nevertheless, the primary mechanism of SET backaction is still the capacitive electromagnetic coupling between the box and SET, and the box remains a mesoscopic device that is sensitive to this backaction. Just as with the SSET-Cooper-pair box system, sensitive measurements of the Coulomb staircase of the normal box can reveal the dynamics of the coupled system, and probe the nature of SET backaction.

The possibility of SET backaction on a single electron box was proposed with the first experiments in the field\cite{Lafarge}, but has proven difficult to quantify. The signature of SET electrical backaction is difficult to separate from simple heating of the sample\cite{Schafer,Krupenin}. The backaction has been measured with very strong coupling between the SET and the box\cite{Heij}, but few measurements exist in systems that are as weakly coupled as the proposed Cooper pair box-SET experiments. In this letter, we present an experimental analysis of an SET weakly coupled to a single electron box. We vary the operating point of the SET, measure the Coulomb staircase of the box, and find the variations in the shift, width, and asymmetry of the staircases to be in agreement with a model that includes backaction caused by the charge state fluctuations of the SET island. These variations in the measured staircases allow us to measure average properties of the noise of the SET.

\begin{figure}
\includegraphics{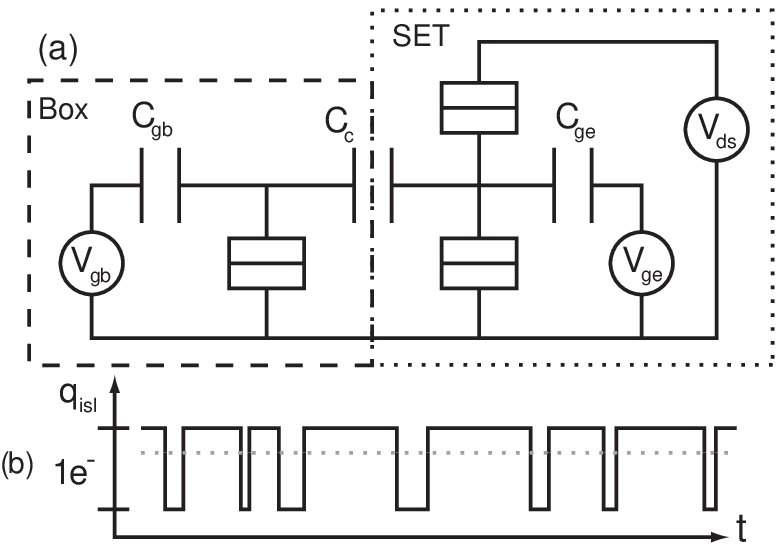}\
\caption{\label{fig:fig1} (a) Circuit diagram of the single electron box (dashed box) capacitively coupled to the SET (dotted box). The normal state tunnel junctions are represented by boxes with a single line through them.\\ (b) Plot of the charge state on the SET island vs. time. The dotted line shows the mean value of the charge on the SET island.}
\end{figure}

The SET (Fig. \ref{fig:fig1}a) consists of a Aluminum island connected through tunnel junctions to two leads (the drain and the source) and capacitively coupled to a third (the gate). An SET is described by its charging energy ($E_{c}=e^{2}/2 C_{\Sigma}$, the energy to add an additional electron to the island), by the tunnelling resistance of the junctions on the drain and the source leads ($R_{j}$), and by the size of the capacitors coupling it to the external control voltage ($C_{ge}$) and to the measured system ($C_{c}$). A high tunnelling resistance ($R_{j}>h/e^{2}$) and large charging energy ($E_{c}>k_{B}T$) suppress the addition of charge to the island by quantum or thermal fluctuations, so the island may be considered confined to a discrete set of charge states. A bias voltage (\Vds) provides the energy necessary for the system to switch between charge states, allowing current to flow from the drain to the source. The amount of current is controlled by the rate of transition between accessible charge states, which is a function of the potential of the island. Thus the SET forms a very sensitive electrometer, where changes in the total charge capacitively coupled to the island modulate the current flowing through the transistor. The SET is operated by fixing the values of the externally applied $V_{ds}$ and $V_{ge}$, and observing variations in the conductance as the charge coupled to the SET from the measured system changes. The point at which $V_{ds}$ and $V_{ge}$ are fixed is termed the operating point; the same measurement can be performed by observing conductance variations about many different operating points.

\begin{figure}
\includegraphics{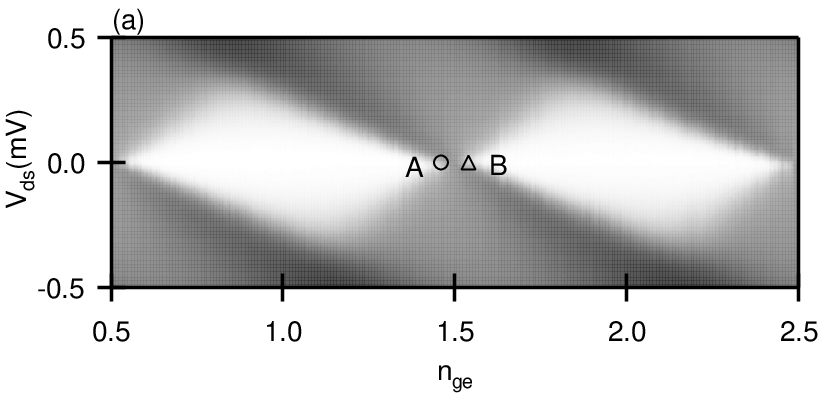}\
\includegraphics{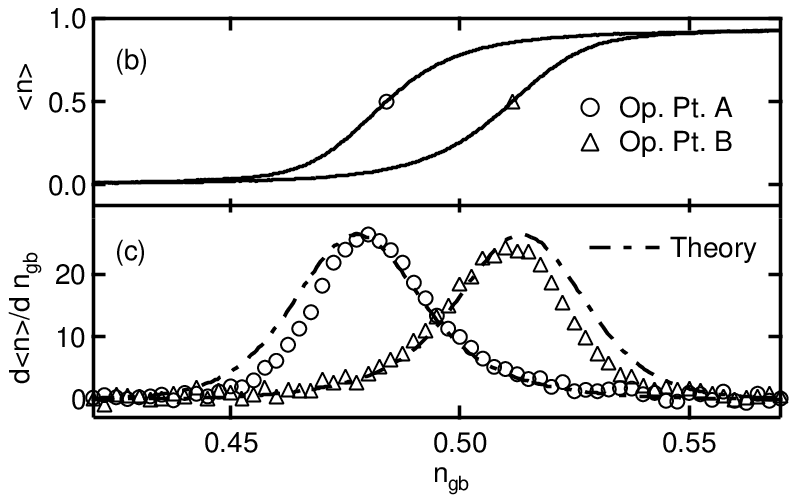}\
\caption{\label{fig:fig2} (a) a plot of the reflected power from the SET as a function of gate(\nge) and drain-source (\Vds) voltage. (b) Coulomb staircases measured at the operating points marked in Fig. \ref{fig:fig2}a as a function of the box gate voltage \ngb. The time averaged number of electrons on the box is measured with a precision of $\pm1\times10^{-3}$ and an accuracy of $\pm2\times10^{-3}$ (c) Derivatives of these Coulomb staircases and of the corresponding Coulomb staircases generated with a sequential tunnelling model with $E_{C_{SET}}/k_{B}=2.3 K$, $E_{C_{box}}/k_{B}=1.6 K$, $R_{j_{SET}}=47 k\Omega$, $R_{j_{box}}=15.4 k\Omega$, and $C_{c}/C_{\Sigma_{SET}}=.048$. The derivative of the Coulomb Staircase is reported with an accuracy of $\pm0.4$}
\end{figure}

The box (Fig. \ref{fig:fig1}a) consists of another island capacitively gated by an external lead (\Vgb) and connected through a tunnel junction to ground. As with the SET, the gate lead controls the potential of the box and changes the relative electrostatic energies of the available charge states. We express the gate voltages for both the box and the electrometer in terms of the number of electrons on the corresponding gate capacitors: $n_{gb}=C_{gb} V_{gb}/e$ and $n_{ge}=C_{ge} V_{ge}/e$. When \ngb\ is raised by 1 electron, the island charge state of minimum energy changes, and a single electron tunnels on to the island to keep it in its ground state. Plotting the time-averaged number of additional electrons on the island as a function of \ngb\ gives the familiar "Coulomb staircase"(Fig. \ref{fig:fig2}b) \cite{Lafarge}. The width of this staircase is normally a function only of the temperature of the sample. In this paper we quantify SET backaction by observing additional variations in the Coulomb staircase that are systematic with SET operating point.

The coupling capacitor ($C_{c}$ in Fig. \ref{fig:fig1}a) couples the potential on the two islands together, allowing the SET to measure the box and also allowing the potential on the SET island to affect the box. The strength of this coupling is expressed either as the fraction of the electrometer charge coupled to the box ($\kappa=\frac{C_{c}}{C_{\Sigma_{SET}}}$) or as the temperature necessary to cause changes in a Coulomb staircase comparable to those caused by backaction ($T_{\kappa}=\frac{\kappa E_{c_{box}}}{k_{B}}$). As the polarization charge on $C_{c}$ changes, the total charge coupled to the SET changes, changing the tunnelling rates in the SET and modulating the current that flows from the drain to the source. The charge on the box is then inferred from the change in current through the SET. $C_{c}$ also couples the charge on the SET island to the box, and in doing so creates the effects that we see as the SET's backaction.

The discrete nature of charge causes two kinds of noise in the SET.  The drain source current flows not as a continuous fluid, but as individual charges, causing an uncertainty in the SET's measurement due to shot noise. In addition to shot noise on the output (the drain-source current), there is also charge noise on the SET input (the gate capacitor) that affects the measured system. Electrons tunnelling on and off the island they cause both the charge state and the potential of the SET island to fluctuate between two values (Fig. \ref{fig:fig1}b). The fluctuating potential on the SET island coupled through \Cc\ is found to be the source of the SET's backaction. Three averaged properties of the fluctuating potential have effects visible on the Coulomb staircase and can be varied with the operating point of the SET.  The mean charge on the SET island varies by as much as 1 electron, and leads to shifts in the position of the Coulomb staircase by as much as $\kappa e$. The RMS magnitude of the charge fluctuations on the SET island broaden the measured Coulomb staircase by an amount that varies with \nge. Finally, the telegraph-noise nature of the charge state fluctuations on the SET island causes the staircases to be asymmetric; the magnitude and direction of that asymmetry varies with the SET's operating point.

A sequential tunnelling model for the full SET-box system accurately recreates both the measurement and the backaction. The tunnelling rates between any two box and SET charge states are calculated as a function of \nge, \ngb, and \Vds (for details, see \cite{DevoretSCT}). The time averaged charge state of the SET-box system corresponds to the steady state of these coupled rates. The current through the transistor is calculated as the product of the time averaged charge on the SET island and the rate at which charge tunnels off the island. This model allows us to replicate the Coulomb staircases taken at various operating points with only the electron temperature as a free parameter.  The elevated temperature of the best fit model steps ($T=27\pm1 mK$ in a fridge at $T=13 mK$) reflected the broadening of the measured steps due to quantum fluctuations of charge \cite{Lehnert}, and is well understood. Theoretical curves also correctly account for higher order effects in the box-SET system. At certain operating points (e.g. \ngeHalf, \VdsZero, \ngbHalf), the SET's backaction is a sensitive function of the state of the box. Changes in the Coulomb staircase measured at such operating points can only be understood by a sequential tunnelling model for the full coupled box-SET system.

Coulomb staircases were measured in a dilution refrigerator at 13 mK, where the available thermal energy was far less than the charging energy of either the SET or the box island. The SET was operated as an RF-SET \cite{SchoelkopfRFSET}, with an LC resonant circuit reflecting an amount of microwave power that varied as the oscillator was damped by the varying conductance of the SET. Staircases were measured by sweeping \ngb\ over a range corresponding to 1/4 e. While the box gate was swept, the SET gate was swept in the opposite direction to cancel the parasitic capacitance of the box lead to the electrometer's island. Before each Coulomb staircase was measured, \nge\ was swept to find the reflected microwave power as a function of charge coupled to the SET island.  Variations in reflected power with \ngb\ were then converted (via this lookup table) to charge on \Cc\ (for a more detailed description, see \cite{Lehnert}).  The measured charge on the box is thus reported from the amount of charge on $C_{ge}$ necessary to cause an equivalent electrometer response.

Backaction effects were found to be very sensitive to variations in \ngb\ and \nge\, and our experiment therefore required that these voltages be set with high precision. Drifts were removed by referencing the steps to a fiducial step every 20 minutes. First, \nge\ was swept at \VdsZero\, and the value of \nge\ that maximized SET conductance was determined as \ngeHalf\ (see Fig. \ref{fig:fig2}a).  Next, a Coulomb staircase was measured with the SET operated at $n_{ge}=.44$, \VdsZero. The value of \ngb\ at the center of this step was determined.  Charge offset noise and 1/f noise drifts add constant offsets to either \nge\ or \ngb; measuring the fiducial step as described here allows us to quantify the change in these offsets on both the box and the SET. Measurements found to contain large charge jumps in \nge\ or \ngb\ were discarded. This procedure allowed measurement of Coulomb staircases with an uncertainty of $1\times10^{-3}e$ in the charge and an uncertainty of $5.5\times10^{-4}e$ in the horizontal position of the steps. The uncertainty in the applied \nge\ was found to be $5\times10^{-3}e$.

\begin{figure}
\includegraphics{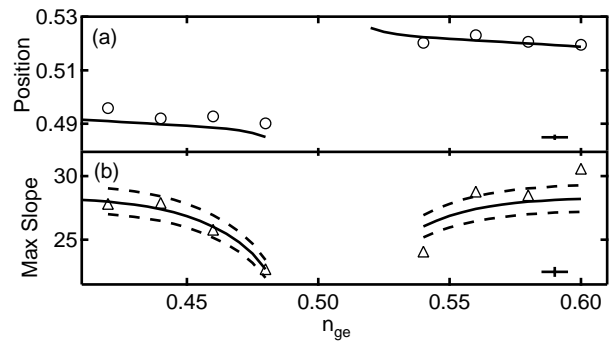}\
\caption{\label{fig:fig3} (a) The horizontal position of the center of the Coulomb staircase for various operating points of the SET.  The model is a solid line, and circles are experimental measurements. No measurements exist near \ngeHalf\ where the electrometer had no gain. Representative error bars are shown in the bottom right of the plot. Horizontal uncertainty reflects the measured instability of the SET operating point due to charge noise.(b) The maximum slope of the staircases measured with the SET at the same series of operating points. Confidence bands show the model curve for $27\pm1 mK$.}
\end{figure}

The differences in Coulomb staircases measured at different operating points allow us to measure average properties of the fluctuating potential of the SET island. Staircases measured at different operating points are shifted in \ngb\ (Fig. \ref{fig:fig2}b). The shift of each staircase is proportional to the mean charge on the SET island. The mean charge on the SET island varies by as much as 1 electron with SET operating point, and the corresponding charge that couples to the box and adds to \ngb\ varies by as much as $\kappa e$. We measure staircase shift by reporting the value of \ngb\ at each step's midpoint, measured relative to the center of a fiducial step (Fig \ref{fig:fig3}a). The sequential tunnelling model accurately recreates these variations in the step position.

The measured Coulomb staircases also exhibit variations in width that change with operating point (Fig. \ref{fig:fig3}b). Three different mechanisms broaden the Coulomb staircase: quantum fluctuations, thermal excitation, and SET backaction. Quantum fluctuations of charge on the box cause broadening, but only away from the center of the step\cite{Lehnert}. Our measurement, which quantifies broadening as the maximum slope at the center of each Coulomb step, is therefore insensitive to quantum broadening.  Thermal excitations of the box also broaden the Coulomb staircase. SET heating varies with operating point, and, for large values of \Vds, can produce a trend in staircase width similar to the effects of backaction. All of our data were taken, however, at \VdsZero, where heating from the SET was negligible. Finally, SET backaction broadens the Coulomb staircase when the charge state fluctuations of the SET island cause the box to switch between charge states. SET backaction broadens staircases by as much as $\kappa e$, and broadens staircases most at operating points where the RMS magnitude of the SET charge state fluctuations is largest. The observed variations in staircase broadening with operating point (Fig. \ref{fig:fig3}b) are fully accounted for with our sequential tunnelling model.

\begin{figure}
\includegraphics{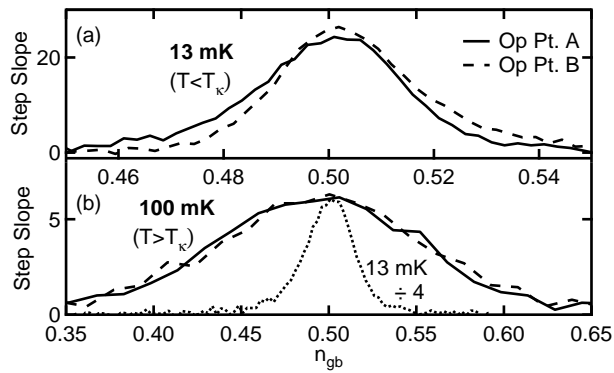}\
\caption{\label{fig:fig4} (a) Derivatives of steps measured at the operating points in Fig. 2a, offset in \ngb\ to eliminate the shift in position of the steps. Note that the tails of the two steps are asymmetric. (b) Steps measured at the same operating points with the sample at 100 mK. The asymmetry is no longer visible. The inset demonstrates the thermal broadening by showing a 1/4 scale curve from the top graph plotted on the x-axis for the bottom graph.}
\end{figure}

The staircases are also asymmetric in a manner that varies predictably with operating point. Each staircase was found to have a longer tail in the direction away from which the staircase was shifted. The asymmetry of the Coulomb staircase is best viewed in the derivative of the steps (Fig. \ref{fig:fig2}b, or with the curves shifted to overlay in Fig. \ref{fig:fig4}a), where it clearly follows the same trend as the model produces. Unfortunately, differentiating our data increased the noise and made it difficult to quantify the asymmetry; qualitatively, however, the model reproduces the experimentally observed trends. The staircase asymmetry is caused by the nature of the charge state fluctuations on the SET island. The potential of the SET island lies preferentially to one side of the mean potential, with infrequent fluctuations far to the other side (Fig. \ref{fig:fig1}b). The staircases are thus broadened asymmetrically in the +\ngb\ and -\ngb\ directions. The preferred charge state, and thus the asymmetry of the measured staircase, is found to switch at \ngeHalf. 

The model also shows good agreement with our data at higher temperatures, where the various effects of the backaction change predictably. At higher temperatures, the mean potential of the SET island still changes with \nge, and thus step shifts are still visible. For $T>T_{\kappa}$, however, the range (in \ngb) of thermal broadening is greater than the range of the backaction broadening or the asymmetry, and neither of these effects are therefore visible (Fig. \ref{fig:fig4}b).

In these experiments we confirm that charge state fluctuations of the SET island are the primary source of SET backaction. We observe the differences in Coulomb staircases measured with the SET biased at a variety of different operating points, and note changes in the shift, width, and asymmetry of the steps that are accurately recreated by a sequential tunnelling model. This confirms that electromagnetic coupling to the fluctuating SET island potential can provide the ultimate lower bound on SET backaction.

This work was supported by the National Security Agency (NSA) and Advanced Research and Development Activity (ARDA) under Army Research Office (ARO) Contract No. DAAD-19-02-1-0045,the NSF ITR program under grant number DMR-0325580, the NSF under grant number DMR-0342157, the W. M. Keck Foundation, and the David and Lucile Packard Foundation.

\end{document}